\documentclass[
reprint,
superscriptaddress,
amsmath,amssymb,
aps,
pra,
twocolumn
]{revtex4-2}

\usepackage{array}
\usepackage{graphicx}
\usepackage{dcolumn}
\usepackage{bm}
\usepackage{xcolor}
\usepackage{colortbl}
\usepackage{ulem} 
\usepackage{mathrsfs}
\usepackage{setspace}
\usepackage{hyperref} 
\usepackage{cleveref} 

\hypersetup{colorlinks = true, 
	    linkcolor = blue, 
	    urlcolor = blue,
            citecolor = blue} 

\newcommand{\red}[1]{\textcolor{red}{#1}}

\begin{document}

\preprint{preparing}

\title{Enhanced photon-pair generation under coherent control}

\author{Ke-Shuang Cui}
	\affiliation{School of Physics and Center for Quantum Sciences, Northeast Normal University, Changchun 130024, China}

\author{Xiao-Jun Zhang}%
	\email{zhangxj037@nenu.edu.cn}
	\affiliation{School of Physics and Center for Quantum Sciences, Northeast Normal University, Changchun 130024, China}

\author{Jin-Hui Wu}%
 	\email{jhwu@nenu.edu.cn}
	\affiliation{School of Physics and Center for Quantum Sciences, Northeast Normal University, Changchun 130024, China}

\date{\today}

\begin{abstract}

The generation of the narrowband strong-correlated biphotons via spontaneous four-wave mixing can be effectively controlled and enhanced by an additional driving field which drives a transition with its upper level being a Rydberg state. We study the properties of the noise of the generated biphotons and show that in the region of weak pumping and low atomic density, a high degree of the photon correlation is maintained with the photon-pair generation rate siginificantly enhanced.

\end{abstract}
\maketitle

\section{Introduction}
Four-wave mixing is the cornerstone of a wide range of fascinating applications, such as one-atom laser \cite{WOS:000185370900037}, entanglement \cite{PhysRevA.69.063803,WOS:000796530400027,WOS:000257888900042}, nonlinear optical amplification \cite{PhysRevA.97.053826}, squeezed light \cite{WOS:000243298700024}, non-Hermitian optical systems \cite{WOS:000495457900007} and microwave-to-optical conversion \cite{PhysRevA.100.012307}.
Especially, the spontaneous four-wave mixing (SFWM) that generates a pair of time-correlated photons in virtue of a third order nonlinearity \cite{PhysRevA.70.031802,WOS:000648815400002,PhysRevA.107.013514,WOS:000262096000014} has driven considerable research efforts. 
Since the detection of the first photon heralds the arrival of the second one, which is employed in further quantum operation, they have been studied in numerous quantum applications, such as quantum communication \cite{WOS:000303149900026,PhysRevResearch.3.013096} and quantum memory \cite{WOS:000279056900047,PhysRevLett.131.150804,PhysRevResearch.2.033155}.

Improving the pair-generation rate has been a longstanding research focus. It was shown that the efficiency can be effectively increased using the nanostructures \cite{WOS:000404332000041,10.1063/1.2814040,Azzini:12} which confine light resonantly, or by introducing other coherent effects to strengthen the light-matter interaction.
For instance, in Ref. \cite{PhysRevA.106.023711} the authors proposed a scheme with double pumping fields to construct a quasidark state \cite{PhysRevA.106.023711,scully:book} that controls the ground population in order to perform the SFWM process near resonance while Raman process is suppressed. However such system supports two sets of SFWM processes that are mixed together, and breaks the one-to-one relation between the Stokes and anti-Stokes photons.
In another example \cite{WOS:001093401100004}, the interaction between the Rydberg atoms is utilized to enhance the generation efficiency via the nonlocal FWM non-linearity \cite{PhysRevLett.105.193603,WOS:000371435000011,PhysRevA.103.043709}. One potential problem is that the Rydberg interaction needs a certain atomic density and the high atomic density usually leads to larger decoherence rates via collisions. This could be fatal to the electromagnetically induced transparency that suppresses the anti-Stokes photons and reduces the degree of the correlation.

Both of the mentioned works use the perturbation theory to model the SFWM process \cite{PhysRevA.82.043814}, where the evolution operator act on  vacuum to produce the Stokes-anti-Stokes two-photon state. The absorption of the anti-Stokes photon and the Raman enhancement of the Stokes photon are phenomenologically included as the imaginary parts of the (complex) wavevector. And the fluctuations of the generated photons are neglected. 

\begin{figure}[b]
\includegraphics[width= 8 cm]{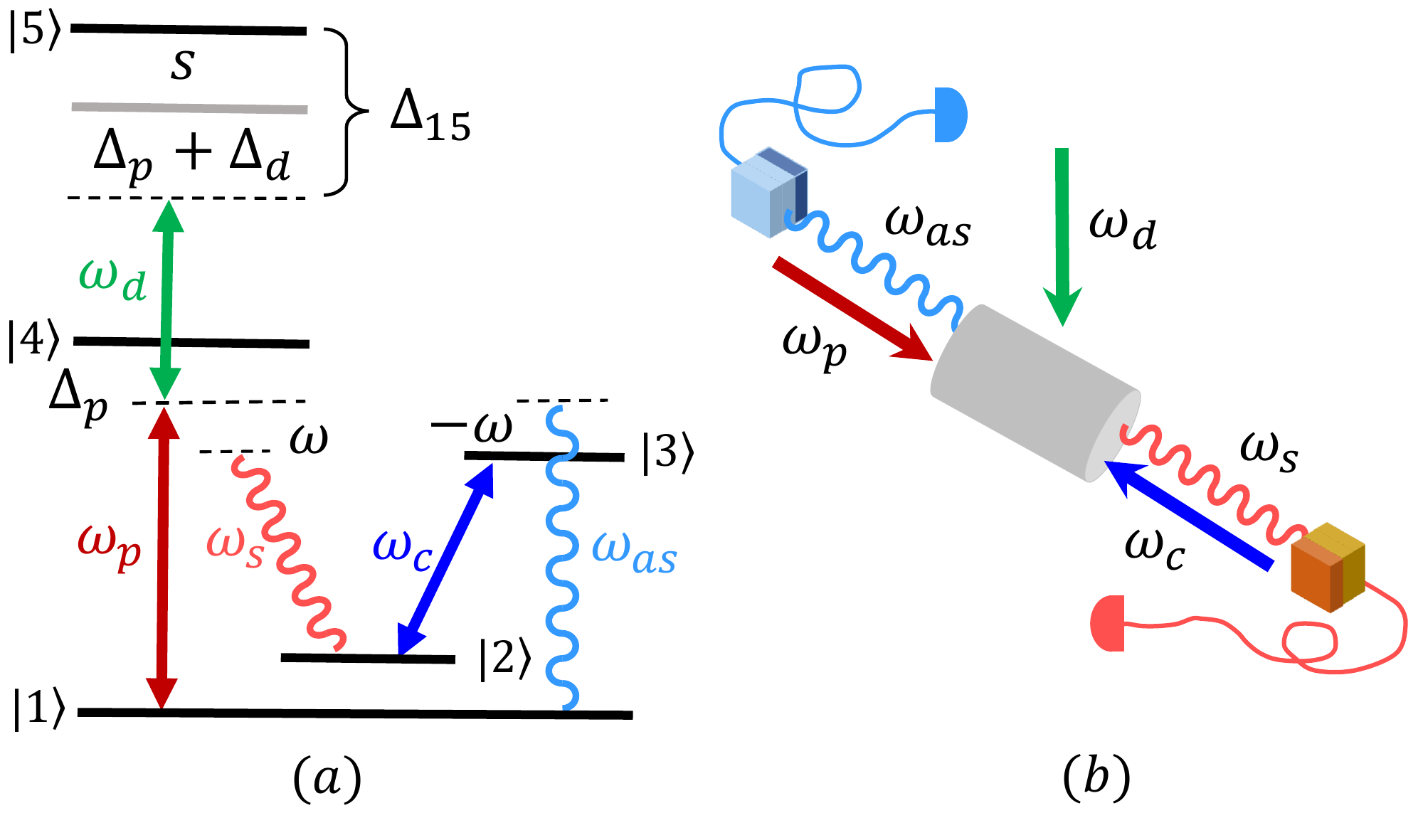}
\caption{\label{fig1}{Energy-level diagram (a) and schematic (b) for spontaneous paired photon generation controlled by a driving field. The sample represented by cylinder is assumed to be a cell of ultra-cold atomic gas $^\text{87}$}Rb. The chosen energy levels are 
$|1\rangle = |5S_{1/2},F=1\rangle$, 
$|2\rangle = |5S_{1/2},F=2\rangle$, 
$|3\rangle = |5P_{1/2},F=1\rangle$, 
$|4\rangle = |5P_{1/2},F=2\rangle$ and 
$|5\rangle$ is a Rydberg level $|60\,S_{1/2}\rangle$.
}
\end{figure}
In this paper, we present a theoretical investigation of a modified model with an additional applied driving field forming a ''ladder`` system with the pumping field, see Fig. \ref{fig1}(a). We use the field operators to model the generated field, and interaction between the fields and the atoms is depicted by a set of differential equations, rather than the evolution operator in perturbation theory. In this frame which is widely used in the related studies
\cite{PhysRevLett.94.183601,PhysRevA.74.023809}, the operators of the fluctuations can be easily accounted in, and allow us to investigate the noise property of the generated photons. Similar to the model discussed in Ref. \cite{WOS:001093401100004}, a Rydberg state is also included, being the upper level of the transition that the driving field couples. However, we only take advantage of the long live time of the level, so that the photons generated by the transitions down from the Rydberg state onto other levels are rare and only one set of SFWM process is supported. As for atomic correlation from the Rydberg-Rydberg interaction, it is negligible due to the weak pumping effect and low Rydberg atomic density that we assumed.

The result shows that the presence of the driving field not only provides an effective way to tune the SFWM process, but also significantly increases the generation rate even without the atomic correlation. Since the efficiency of the Raman gain and the absorption increase much slower than that of the SFWM nonlinearity, the noises of the generated photons remain at the safe level and the photon correlation is considerably high.

Our paper is organized as follows. In Sec. \ref{sec2} we describe our system and derive the dynamical equations for the field operators, the coefficients of which correspond to the effects of Raman enhancement, linear absorption, four-wave mixing nonlinearity, and fluctuations. The properties of the coefficient are examined as well. In Sec. \ref{sec3}, the influence of the driving field on the generation rate and the correlation between the Stokes and anti-Stokes photons are investigated. We conclude in Sec. \ref{sec5} and the full set of equations for the elements of the density matrix, the expression of coefficients, and elements of the diffusion coefficient are presented in Appendix \ref{appA}, \ref{appB} and \ref{appC}.

\section{\label{sec2}Model and equation}
Let us consider a group of ultra-cold five-level atoms, as shown in Fig. \ref{fig1} (a), in the magneto-optical trap where the thermal motion is effectively suppressed so that we do not need to include Doppler broadenings in the following calculation. The pumping field at frequency $\omega_p$ drives the transition $|1\rangle\leftrightarrow|4\rangle$, while the coupling field at frequency $\omega_c$,  travels in the opposite direction of the pumping (see Fig. \ref{fig1}(b)), and drives $|2\rangle\leftrightarrow|3\rangle$.  
The corresponding Rabi frequency of the applied fields are $\Omega_\alpha = \mu_{mn}E_\alpha^+/ 2\hbar$
 $(\alpha \in \{c,p\})$ where $\mu_{mn}$ is the dipole moment of the corresponding transition and $E_\alpha^+$ is the positive-frequency part of the electric field.
In virtual of the third-order nonlinearity, a pair of photons is generated spontaneously from the transition $|2\rangle\leftrightarrow|4\rangle$ and $|1\rangle\leftrightarrow|3\rangle$, and we refer to them as the Stokes photon (the one at freqeuncy $\omega_s$) and anti-Stokes photon (at $\omega_{as}$).
To suppress the generation of impurity photons from the spontaneous decays and the Raman process, the pumping field is tuned largely off the resonance, \text{i.e.} the detuning of the pumping $\Delta_p = \omega_{41} - \omega_p$ is much larger than the decoherence rate $\gamma_{41}$. In this way, the Stokes photons are also far away from their resonance and the corresponding absorptions are effectively limited.
The coupling field resonates with its transition and creates a window due electromagnetically induced transparency (EIT) \cite{PhysRevLett.66.2593,RevModPhys.77.633}, and allows the anti-Stokes photons to pass through the atomic medium. Traditionally, only the pumping and coupling fields are applied.

In our system, an additional driving field with Rabi frequency $\Omega_d$ and frequency $\omega_d$ is introduced to couple the transition between $|4\rangle$ and an even higher atomic level $|5\rangle$ which is assumed to be a Rydberg state. The detuning of the driving field is $\Delta_d = \omega_{54}-\omega_d$.
We use \red{$\omega$} denoting the frequency difference between the pumping field and Stokes photon, $\omega = \omega_p - \omega_{s}$, then the interaction Hamiltonian reads
%
%
$\hat {V}=- \hbar \sum_{i=1}^N\big[
\Delta_p\hat{\sigma}_{44}^{[i]} + (\Delta _{p} + \Delta_{d})\hat{\sigma}_{55}^{[i]} -\omega \hat{\sigma}_{22}^{[i]}
+ \omega \hat{\sigma}_{33}^{[i]}
+g_{as}\hat {a}_{as}\hat{\sigma} _{31}^{[i]}
+g_{s}\hat {a}_{s}\hat {\sigma} _{42}^{[i]}+\Omega _{c}\hat  {\sigma} _{32}^{[i]}+\Omega _{p}\hat {\sigma} _{41}^{[i]}+ \Omega _{d}\hat  {\sigma} _{54}^{[i]}+ \text{H.c.}\big]+\frac{1}{2}\sum_{i,j=1}^N\big[\hat{\sigma}_{55}^{[i]}(C_6/|\mathbf{r}_i - \mathbf{r}_j|^6)\hat{\sigma}_{55}^{[j]}\big].$
$N$ is the number of atoms in the sample. $\hat{\sigma}_{mn}^{[i]} =|m\rangle_{ii}\langle n|$ is the transition operator for the $i^\text{th}$ atom. 
Note that the conservation of energy in the four-wave mixing process is already applied in the Hamiltonian. This means the frequency difference between the coupling field and anti-Stokes photon are $\omega_c - \omega_{as} = -\omega$. 

%
For the generated photons, the coupling constant $g_{\beta}=(\mu_{mn}E^+_{\beta})/(2\hbar)$ ($\beta\in \{s,as\}$) and $E^+_{\beta}=\sqrt{(\hbar \omega_{m})/(2\epsilon_{0}V)}$ is the electric field of a single photon. The last term in $\hat{V}$ depicts the interaction between the Rydberg atoms, where $|\mathbf{r}_i - \mathbf{r}_j|$ is the distance between the $i^\text{th}$ and $j^\text{th}$ atom, and $C_6$ is the strength coefficient. 

\begin{figure}[t]
\includegraphics[width= 7.0 cm]{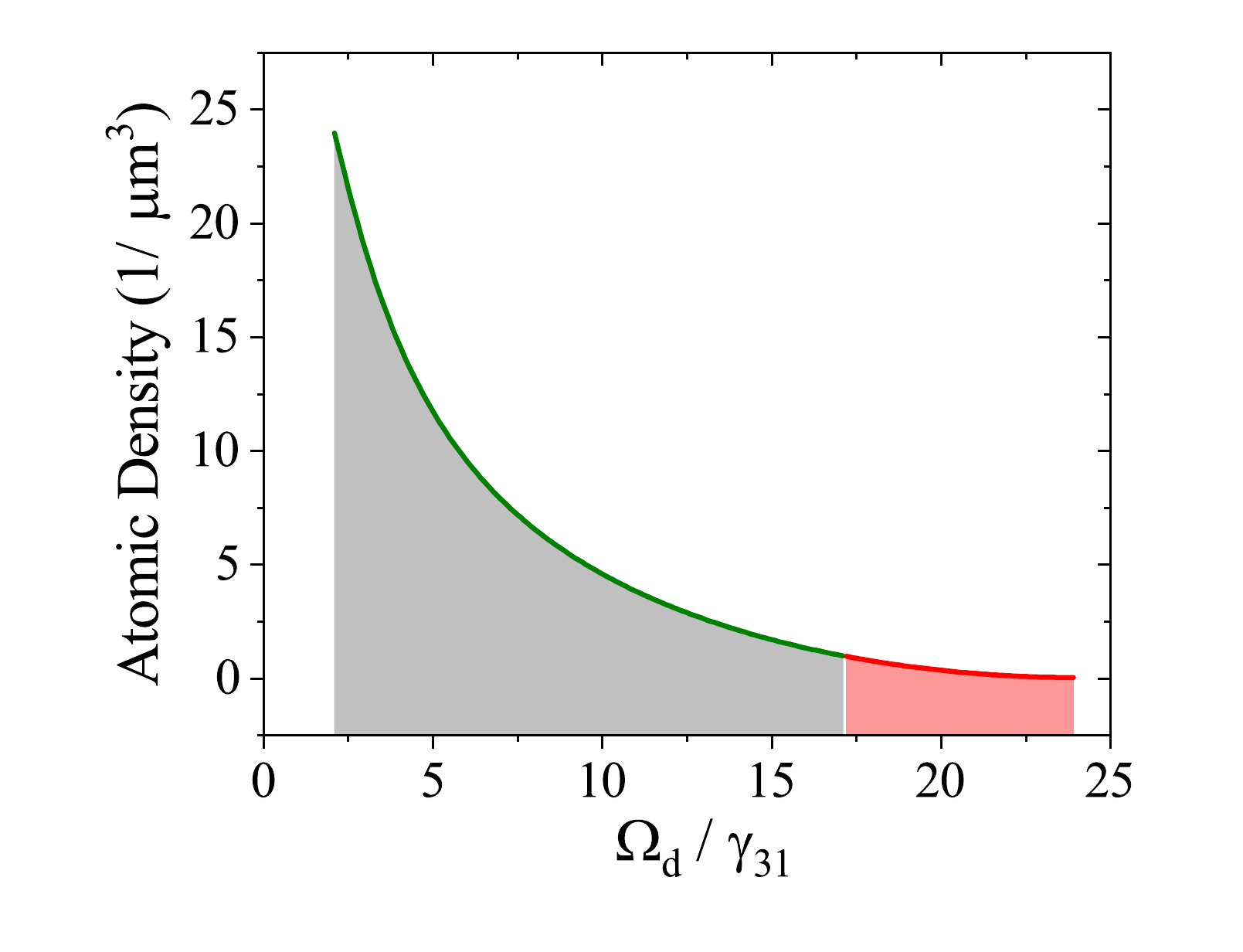}
\caption{\label{figA}
The maximal atomic density allowed for neglecting the atomic correlation under different driving Rabi frequencies. 
The gray area and pink area is separated by a white vertical line which indicates the point at $\Omega_d = 17\, \gamma_{31}$ with the corresponding atomic density being $N/V = 1\, \mu m^{-3}$. Here 
$\Delta_{p}=24\,\gamma_{31}$,
$\Delta_{c} = 0$,
$\Delta_{15} = 24\,\gamma_{31}$
$\Omega_{p}=1.2\,\gamma_{13}$,
$\Omega_{c}=3.0\,\gamma_{13}$,
$\gamma_{12}=10^{-3}\,\gamma_{13}$,
$\Gamma_{42}=\Gamma_{41}=\Gamma_{32}=\Gamma_{31}=\gamma_{31}$,
$\Gamma_{5,4}=\Gamma_{5,3} = 10^{-3}\gamma_{13}$.
}
\end{figure}

A Rydberg atom prevents the other atoms within a distance of $R_b$ from excited up a Rydberg level. $R_b$ normally referred to as the blockade radius 
\cite{RevModPhys.82.2313} is $R_b = (C_6/\delta_\text{EIT})^{1/6}$ with $\delta_\text{EIT}$ being the linewidth of EIT transmission spectrum. In our case with $\Delta_p \gg \gamma_{41}$, $\delta_\text{EIT} = |\Omega_d|^2 / \Delta_p$. 
Besides the Rydberg blockade, the long-range interaction induces correlation ($\langle\hat{\sigma}_{\alpha\beta}^{[i]}\hat{\sigma}_{\mu\nu}^{[j]}\rangle$) between the states of the two Rydberg atoms, consequently, the nonlocal nonlinearity of the medium \cite{WOS:000371435000011,PhysRevLett.105.193603,WOS:000863490000001}. Such nonlinearity can modify the optical responses of the medium. However, If the Rydberg population is low enough, the modification is negligible. Specifically, if the probability of finding a Rydberg atom in the sphere of $3R_b$ is less than unity, that is
\begin{equation}\label{condition}
\frac{4}{3}\pi (3 R_b)^3 \frac{N}{V} \langle \hat{\sigma}_{55} \rangle \leq 1,
\end{equation}
with $V$ being the volume of the atomic cell represented by the cylinder in Fig. \ref{fig1}(b), then we can keep the atomic correlation out of our consideration. Here $\langle \hat{\sigma}_{55} \rangle$ is the population on $|5\rangle$, and it can be obtained from the dynamical equation related to $\hat{V}$ (see the latter discussions).
Eq. (\ref{condition}) can be satisfied if we set a reasonably low atomic density and weak coupling from $\Omega_p$ and $\Omega_d$.
In Fig. \ref{figA} we show the relation between the $N/V$ and $\Omega_d$ when the equal sign is taken in the above condition. With other parameters fixed, the boundary atomic density is reduced as the driving field becomes stronger. For every point below the curve, the atomic density is safe for neglecting the atomic correlation and our following calculation is valid. 

\begin{figure}[t]
\includegraphics[width= 8.0 cm]{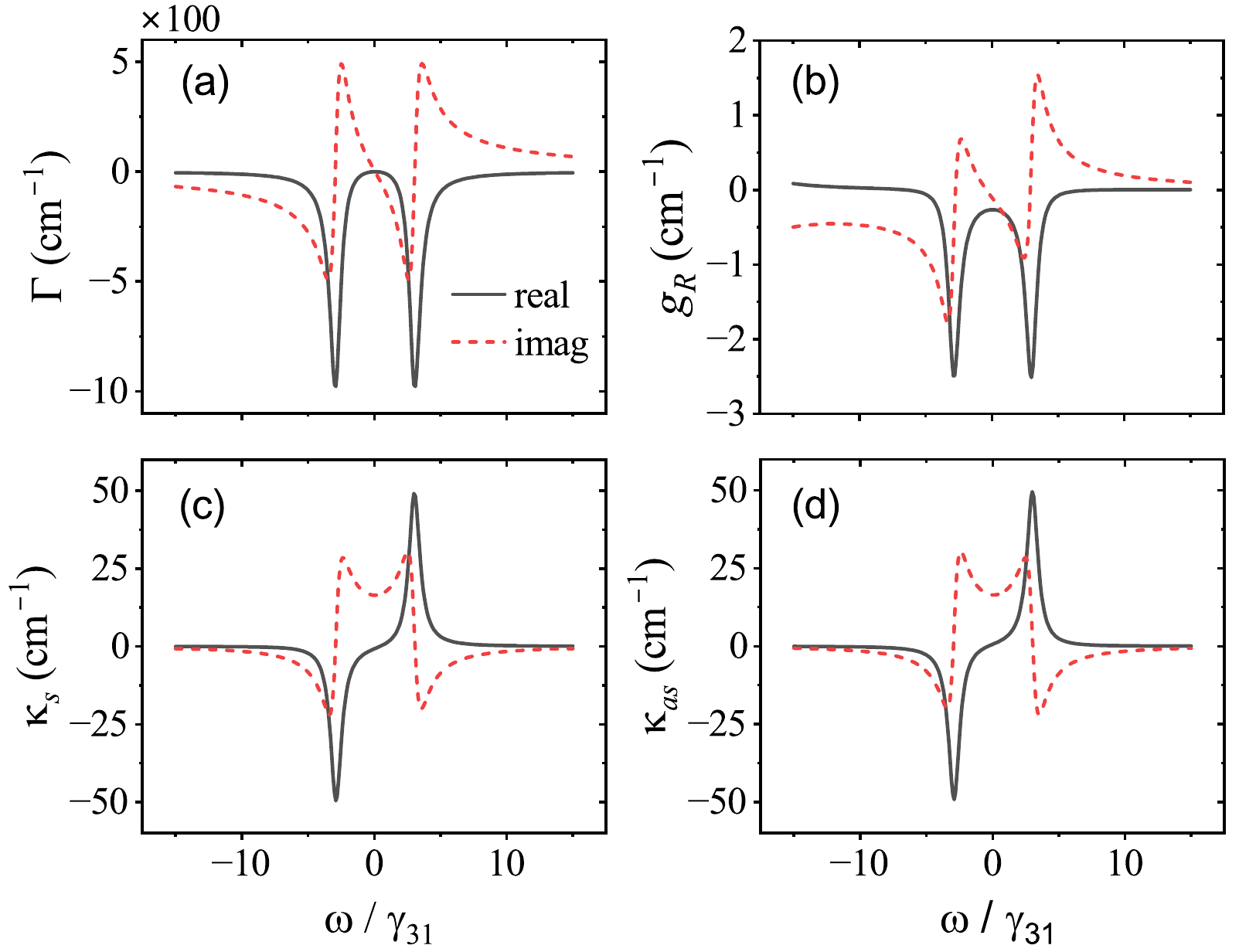}
\caption{\label{figB}
The coefficient of absorption $\Gamma$ (a), Raman enhancement $g_R$ (b), Stokes and anti-Stokes nonlinear coefficients $\kappa_s$ (c) and $\kappa_{as}$ (d) plotted as the functions of the double photon detuning $\omega$. Here $\Omega_{d}=1.2\,\gamma_{13}$,
$N = 1.0\times10^{12}\, \text{ cm}^{-3}$,
$\sigma = 10^{-9}\, \text{ cm}^{2}$
and 
$L = 0.01 \text{ cm}$.
The other parameters are identical with that in Fig. \ref{figA}.
}
\end{figure}

\begin{figure}[t]
\includegraphics[width= 7.5 cm]{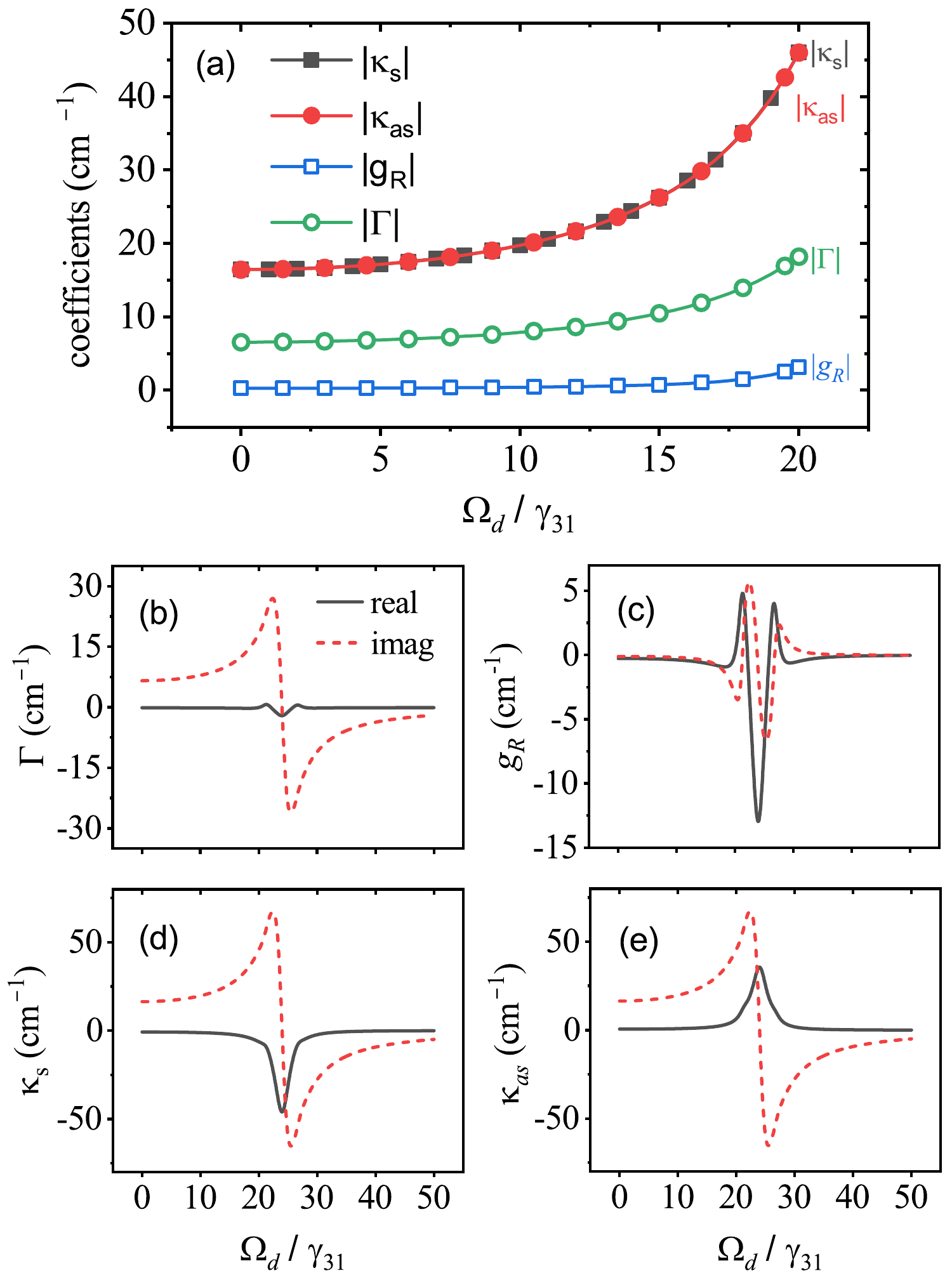}
\caption{\label{figC}The relation between the four coefficients and the driving Rabi frequency.
(a) $|\Gamma|$, $|g_R|$, $|\kappa_s|$ and $|\kappa_{as}|$ under different driving Rabi frequencies.
The real and imaginary parts of the coefficients are shown in (b)-(e) where a wider range of $\Omega_d$ is adopted.
Here $\omega = 0$, and other parameters are identical with that in Fig. \ref{figB}.
}
\end{figure}

Under such circumstances, The Rydberg atoms are far from each other, and it is reasonable to use mean-field treatment (essentially $\langle\hat{\sigma}_{\alpha\beta}^{[i]}\hat{\sigma}_{\mu\nu}^{[j]}\rangle \simeq \langle\hat{\sigma}_{\alpha\beta}^{[i]}\rangle\langle\hat{\sigma}_{\mu\nu}^{[j]}\rangle$) to replace the Rydberg-Rydberg interaction by an effect shift on the level $|5\rangle$, see Fig. \ref{fig1}(a).
The amount of the shift \cite{WOS:000461051100005,PhysRevA.93.022709} is determined by the overall effect of the influence from other nearby excited atoms: $s \simeq \frac{N}{V}\int_{R_b}^{\infty}\frac{C_6}{r^6}\langle\hat{\sigma}_{55}\rangle 4\pi r^2dr$. Thus the effective double-photon detuning of the transition $|1\rangle\leftrightarrow|5\rangle$ is denoted by $\Delta _{15} = s+\Delta_p + \Delta_d$. In this way, we can neglect the Rydberg-Rydberg interaction in the Hamiltonian and the collective slowly varying atomic operators $\hat  {\sigma}_{mn}=N^{-1} \sum _{i=1}^N |m\rangle_{ii}\langle n|$ evolves according to
the Heisenberg-Langevin equations
\begin{equation}\label{sigma.mn}
 \frac{\partial }{\partial  t} \hat  {\sigma}_{mn} =\frac{i}{\hbar}[\hat{V},\hat{\sigma}_{mn}] +\hat{F}_{mn}(z,t).
\end{equation}
Where $\hat{F}_{mn}$ is the collective atomic $\delta$-correlated Langevin noise operators and subject to 
\begin{equation}
\begin{split}
    \langle\hat{F}_{mn}(z,t)\hat{F}_{m'n'}&(z',t')\rangle=
    \frac{L}{N}\mathcal{D}_{mn,m'n'}\delta(t-t')\delta(z-z'),
\end{split}
\end{equation}
with $L$ being the length of the sample along the propagation of the field. 
Eq. (\ref{sigma.mn}) corresponds to a set of differential equations which are presented in Appendix \ref{appA}. 
The Langevin diffusion coefficient $\mathcal{D}_{mn,m'n'}$ is determined by the generalized fluctuation-dissipation theorem and Einstein relation, and we list them in Appendix \ref{appC}. At the steady state, they are reduced to algebraic equations and we can solve them pertubatively with respect to the generated field $\hat{a}_s$, $\hat{a}_{as}$ and the noise operator $\hat{F}_{mn}$. The zero-order solution $\langle\sigma_{ij}^{(0)}\rangle$ corresponds to the case without the generated field. For example, the Rydberg population $\langle\sigma_{55}\rangle$ in the condition (\ref{condition}) and the expression of $s$, is estimated by its zero-order value $\langle\sigma_{55}^{(0)}\rangle$.

Investigating the properties of, \textit{e.g.} the generation rate of Stoke photons relies on simulating its coupling effect with the anti-Stokes field (nonlinearity of four-wave mixing) and the Raman gain it experienced due to the pumping field. To this end, we need to obtain the first-order solution of $\sigma_{24}$ to find the corresponding polarization $P =2  N \mu_{mn} \sigma_{24}^{(1)}$. Then the generation and propagation of the Stokes photons is subject to the Maxwell equation whose Fourier transform reads \cite {PhysRevA.75.033814}:
\begin{subequations}
\begin{equation}\label{eq.s}
\frac{\partial \hat  {a} _{s} }{\partial  z} +g_{R}\hat{a}_{s}+\kappa_{s}\hat{a}_{as}^{\dagger}=\sum_{\alpha}\xi^{s}_{\alpha}\hat{F}_{\alpha},
\end{equation}
where $g_{R}(\omega)$ depicting the effect of Raman gain corresponds to the part of $\sigma_{24}^{(1)}$ that is proportional to $\hat{a}_s$. And $\kappa_{s}(\omega)$ is the coefficient of the interaction between the Stokes and anti-Stokes photons, it is derived from the part of $\sigma_{24}^{(1)}$ proportional to $\hat{a}_{as}^{\dagger}$. The rest of the $\sigma_{24}^{(1)}$ is composed of six parts that are respectively proportional to the noise operators 
$\hat{F}_{21}$, $\hat{F}_{31}$, $\hat{F}_{24}$, $\hat{F}_{34}$, $\hat{F}_{25}$ and $\hat{F}_{35}$. Their contributions to the evolution of the Stokes photons are collected in the right-hand side of Eq. (\ref{eq.s}) with summation with respect to $\alpha$ is taken over $\alpha \in \{21,31,24,34,25,35\}$. 
A similar equation can be written for the anti-Stokes photons
\begin{equation}\label{eq.as}
\frac{\partial \hat  {a} _{as}^{\dagger} }{\partial  z} +\Gamma_{as}\hat{a}_{as}^{\dagger}+\kappa_{as}\hat{a}_{s}^{\dagger}=\sum_{\alpha}\xi^{as}_{\alpha}\hat{F}_{\alpha}.
\end{equation}
\end{subequations}
Here $\Gamma_{as}(\omega)$ is the coefficient linear absorption. And $\kappa_{as}(\omega)$ is the coupling coefficient.  The coefficients appear in Eq. (\ref{eq.s}) and (\ref{eq.as}) are listed in Appendix \ref{appB}.

In Fig. \ref{figB} we show the value of $\Gamma$, $g_R$, $\kappa_s$ and $\kappa_{as}$ for different $\omega$. As we can see that $\Gamma$ has a typical profile of EIT  created by the resonant coupling field. Similar patterns are manifested in the other three coefficients as well. For example, when the absorption of the anti-Stoke photon is large (minimum of Re $\Gamma$), the Raman enhancement is severe as well. The valuable region is around the double-photon resonance ($\omega \sim 0$) where the absorption and Raman enhancement is limited, and the FWM strength ($|\kappa_s|, |\kappa_{as}|$) is still considerable.

The effect of the driving field can be examined by fixing the other parameters and only varying it alone. $|\kappa_s|$ and $|\kappa_{as}|$, as shown in Fig. \ref{figC}(a), always take the same value for a given value of $\Omega_d$. They increase as $\Omega_d$ increases. This implies that the efficiency of the paired-photon generation should be increased as well. However, the absorption and Raman enhancement are enhanced as well. Fortunately, they grow much slower.

Such monotonic increments of the four coefficient with respect to $\Omega_d$ are better understood if we plot them over a larger range of $\Omega_d$, which are shown in Fig. \ref{figC}(b)-(e). The stronger driven field dynamically shifts the energy level of $|2\rangle$, and makes the coefficients exhibit an approximate Lorentz lineshape as resonances at a particular value of $\Omega_d$, in our case they happen at $\Omega_d = 24 \gamma_{31}$. The increment we show in Fig. \ref{figC}(a) happens on the left-hand side of the resonance. Note that increasing $\Omega_d$ causes the population on $|5\rangle$ to increase as well. This might lead to the situation that the influence of the atomic correlation due to the Rydberg interaction is not negligible anymore. Considering that these four coefficients shown in Fig. \ref{figC} (b)-(e) are proportional to the atomic density, One can always reduce the atomic density to make the atomic correlation reasonably small. This simply scales down the values shown in the figures but the lineshape remains unchanged, and so does this resonance property.

\section{\label{sec3}Generation rates and the correlations of the biphotons}
The generation rate can be obtained from the solution of the Eqs. (\ref{eq.s}) and (\ref{eq.as}) relates to an ``evolution'' operator 
\begin{equation}
\mathcal{M} = \exp\left(
\begin{bmatrix}
g_R&\kappa_s\\
\kappa_{as}&\Gamma_{as}
\end{bmatrix}
\right) = 
\begin{bmatrix}
a&b\\
c&d
\end{bmatrix}.
\end{equation}
In the scheme shown in Fig. \ref{fig1}(b) coupling and pumping fields travel in opposite directions. According to the phase matching condition ($\mathbf{k}_c + \mathbf{k}_p = \mathbf{k}_s + \mathbf{k}_{as}$), the Stokes photon also travels along the opposite direction of the anti-Stokes photon. Thus, $[\hat{a}_{s}(0), \hat{a}^{\dagger}_{as}(L)]^T$ should be treated as the boundary condition when solving the differential equations (\ref{eq.s}) and (\ref{eq.as}), the solution of $[\hat{a}_{s}(L), \hat{a}^{\dagger}_{as}(0)]^T$ are
\begin{equation}\label{solution}
\begin{bmatrix} \hat{a}_{s}(L) \\ \hat{a}_{as}^{\dagger}(0) \end{bmatrix}
=\begin{bmatrix}
A & B \\
C & D 
\end{bmatrix}
\begin{bmatrix} 
\hat{a}_{s}(0) \\ \hat{a}_{as}^{\dagger}(L) 
\end{bmatrix}
+\sum_{\alpha}\int_{0}^{L}dz 
\begin{bmatrix} P_{\alpha} \\ Q_{\alpha}
\end{bmatrix}
\hat{F}_{\alpha}.
\end{equation}
Where
\begin{equation}
\begin{bmatrix} 
A & B \\
C & D 
\end{bmatrix}
=\begin{bmatrix} 
a-bc/d & b/d \\
-c/d & 1/d 
\end{bmatrix},
\end{equation}
\begin{equation}
\begin{bmatrix} P_{\alpha} \\ Q_{\alpha}\end{bmatrix}
=\begin{bmatrix} 
1 & -b/d \\
0 & -1/d 
\end{bmatrix}
e^{\mathcal{M}(z-L)}
\begin{bmatrix} 
\xi^{s}_{\alpha} \\ \xi^{as}_{\alpha}
\end{bmatrix}.
\end{equation}
Then the generation rate of the Stokes photon can be written as
\begin{equation}
R_{s}=\frac{c}{L}\langle \hat{a}_{s}^{\dagger}(L,t)\hat{a}_{s}(L,t)\rangle.
\end{equation}
Where $\hat{a}_{s}(L,t)$ is the Fourier transform of the $\hat{a}_{s}(L)$ [see Eq. (\ref{solution})]. The frequency dependence of the latter on \red{$\omega$} is not explicitly shown.
Using the commutation relation $[a_{j}(z_{j},\omega),a_{j}^{\dagger}(z_{j},\omega ')]=L/(2\pi c)\delta(\omega-\omega')$ with $j\in\{s,as\}$
and the correlation $\langle\hat{F}_{\alpha}^{*}(\omega,z)\hat{F}_{\alpha}(\omega',z')\rangle=L/(2\pi N)\mathcal{D}_{\alpha,\alpha^*}\delta(\omega-\omega')\delta(z-z')$,
we can rewrite the generation rates as 
$R_{s}=\int \frac{d\omega}{2\pi}\tilde{R}_s(\omega)$,
with spectral generation rates $\tilde{R}_{s}$ based on the solution (\ref{solution}) is
$\tilde{R}_{s}(\omega)=|B(\omega)|^{2}+\int_{0}^{L} \sum_{\alpha,\alpha^{*}} P_{\alpha}^{*}\mathcal{D}_{\alpha^*,\alpha}P_{\alpha}dz$, 
Then the generation rate can be written as: 
\begin{equation}
R_s = R_s^{D}+R_s^{F}, 
\end{equation}
with the deterministic part
\begin{subequations}
\begin{equation}
R_s^{D} = \int \frac{d\omega}{2\pi}|B(\omega)|^2.
\end{equation}
And the contribution of the fluctuations is
\begin{equation}
R_s^{F} = \int \frac{d\omega}{2\pi}\int_{0}^{L} \sum_{\alpha,\alpha^{*}} P_{\alpha}^{*}\mathcal{D}_{\alpha^*,\alpha}P_{\alpha}dz.
\end{equation}
\end{subequations}
The summation with respect to $\alpha^*$ in above equation originates from  $\hat{a}_{s}^\dagger(L,\omega)$ and $\hat{a}_{as}(0,\omega)$, and it is taken over $\alpha^*\in\{12,13,42,43,52,53\}$. For the generation of the anti-Stoke photons, the generation rate is 
\begin{equation}
R_{as}=R_{as}^{D}+R_{as}^{F},
\end{equation}
with
\begin{subequations}
\begin{equation}
R_{as}^{D} = \int \frac{d\omega}{2\pi}|C(\omega)|^2,
\end{equation}
\begin{equation}
R_{as}^{F} = \int \frac{d\omega}{2\pi}\int_{0}^{L} 
\sum_{\alpha,\alpha^{*}} Q_{\alpha}\mathcal{D}_{\alpha,\alpha^*}Q_{\alpha}^{*}dz.
\end{equation}
\end{subequations}

\begin{figure}[t]
\includegraphics[width= 8.0 cm]{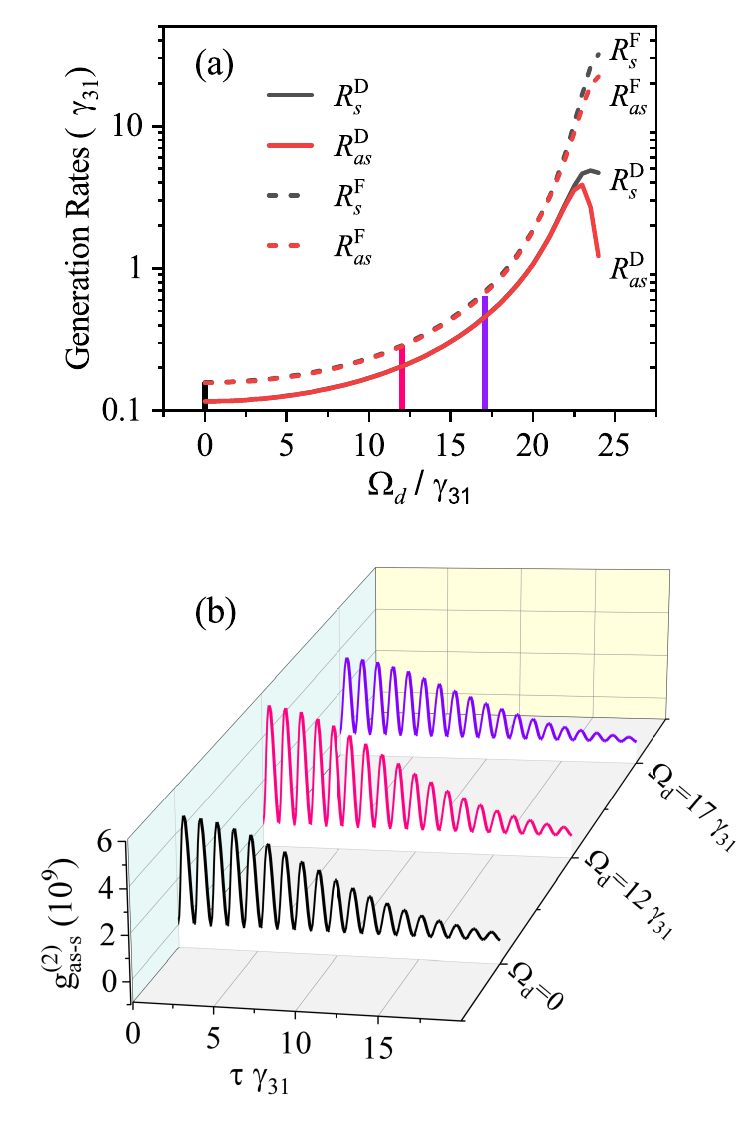}
\caption{\label{figD}
(a): Generation rate plotted against the driving Rabi frequency. (b): The normalized cross-correlation function for three particular values of $\Omega_d$ which are indicated by the vertical line in (a). The parameters are chosen to be the same as that in Fig. \ref{figB}.
}
\end{figure}

Since the coefficients of four-wave mixing increase with respect to $\Omega_d$, one would naturally expect that the generation rates should be enhanced when stronger driving field is applied. Such behaviour is shown in Fig. \ref{figD}(a). As we can see that
$R_{s}^{D}$ and $R_{as}^{D}$, the generation rates without fluctuations, remain identical until the stronger driven field is applied. However, larger coupling fields also lead to the relatively large Rydberg population and considerable atomic correlation. Suggested by Fig. \ref{figA}, our theory is valid for $\Omega_d$ less than $17~\gamma_{31}$, and the data corresponding to the larger driving Rabi frequency is still informative, but inaccurate. 

The degree of the correlation between the generated photons can be represented by Glauber's second-order correlation function which is defined as 
\begin{equation}
\begin{split}
G_{as,s}^{(2)}(\tau) &= \langle
\hat{a}^{\dagger}_{as}(0,t)
\hat{a}^{\dagger}_{s}(L,t+\tau)
\hat{a}_{s}(L,t+\tau)
\hat{a}_{as}(0,t)\rangle
\\ &=R_s R_{as}+|\Phi_{as,s}(\tau)|^2.
\end{split}
\end{equation}
With
\begin{equation}
\begin{split}
&\Phi_{as,s}(\tau) = \frac{L}{c}\mathscr{F}
\bigg(\sum_{\alpha,\alpha^*}\int_0^L dz P_{\alpha}D_{\alpha,\alpha^*}Q_{\alpha}^*+AC^*\bigg).
\end{split}
\end{equation}
Here $\mathscr{F}$ stands for the Fourier transformation. The quantity that we are interested in is the normalized correlation function $g_{as,s}^{(2)} = 1 + |\Phi_{as,s}(\tau)|^2/(R_sR_{as})$ which is the degree of correlation between Stokes and anti-Stokes photons. The condition for biphoton generation is $g_{as,s}^{(2)}\gg 1$. 

We choose three different cases with $\Omega_d = 0$ (transitional four-level system), $12 \, \gamma_{31}$, and $17 \, \gamma_{31}$ (boundary where the theory applies) to show in Fig. \ref{figD}(b) the variation of $g_{as,s}^{(2)}$ with respect to the driving field. 
The correlation function oscillates at the frequency of the $\Omega_c$ which is a typically effect of EIT-based interaction. 
The large values of $g_{as,s}^{(2)}$ appear in the case of weaker driving field ($\Omega_d \to 0$). And
the trade-off between the generation rate and the degree of correlation is clear:  Overall speaking, $g_{as,s}^{(2)}$ is reduced as $\Omega_d$ increases. 
Within the scope that our calculations remains reasonable, the degree of the correlation between the Stokes and anti-Stokes photons is considerably high.

\begin{figure}[t]
\includegraphics[width= 8.0 cm]{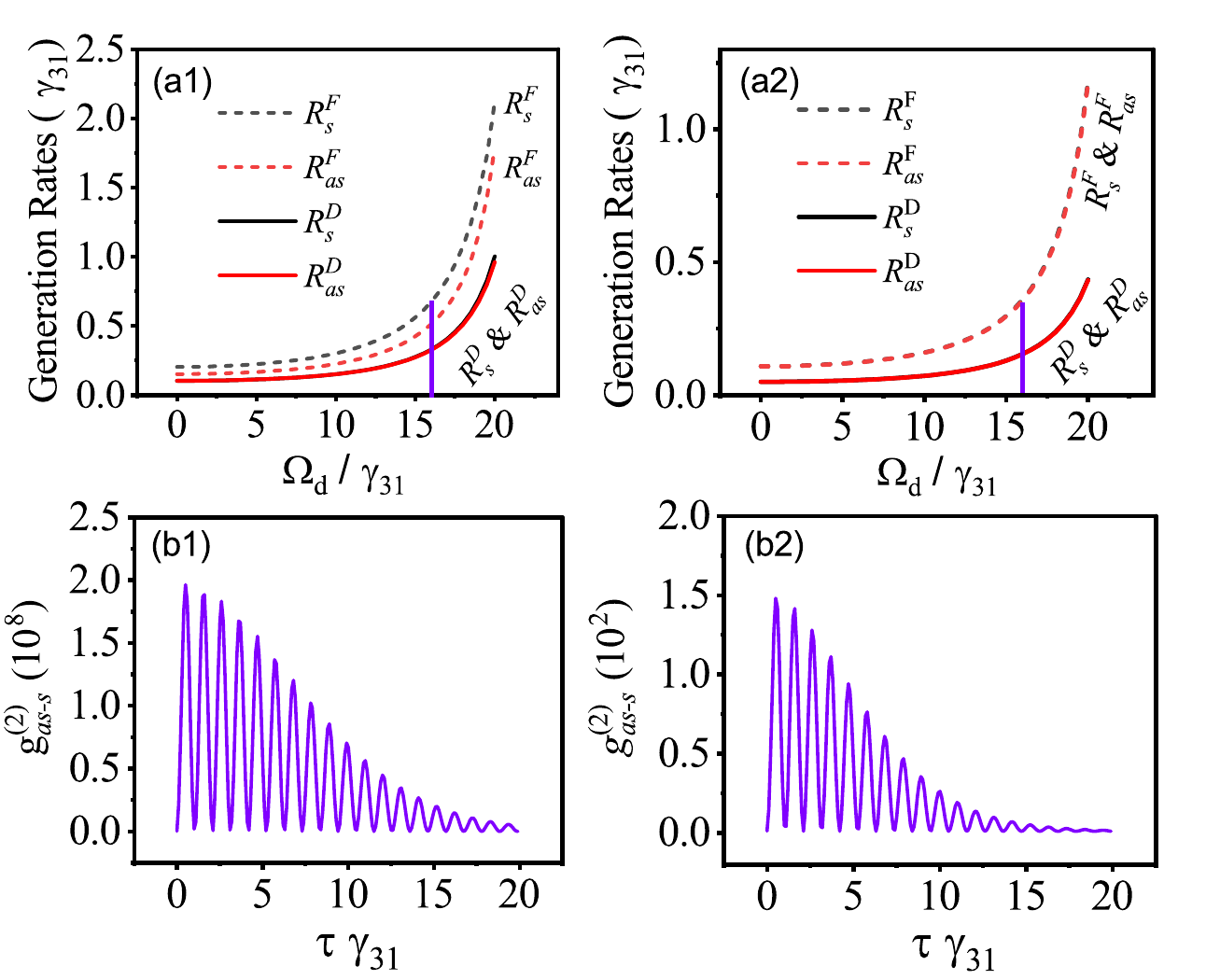}
\caption{\label{figF}
Generation rates plotted as functions of the driving Rabi frequency with $\gamma_{21} = 0.1\, \gamma_{31}$ and $N/V = 1\, \mu m^{-3}$ (a1). The vertical line corresponds to $\Omega_d = 17 \gamma_{31}$ at which the degree of the photon correlation function is shown in (b1). In (a2) and (b2), $\gamma_{21} = 10^{-3}\, \gamma_{31}$ and $N/V = 0.5\, \mu m^{-3}$. The parameters are chosen to be the same as that in Fig. \ref{figB}.
}
\end{figure}

For relatively weak driving, it appears that $R_{s}^{F} = R_{as}^{F}$. This is a result of small decoherence rate between the ground levels ($\gamma_{21} = 10^{-3}\,\gamma_{31}$ in Fig. \ref{figD}) that we assumed. If a hotter atomic gas is used as the sample, then $\gamma_{21}$ is increased. This leads to imperfect EIT, meaning the severe absorption on the anti-Stokes photons. In turn, the generation rate due to fluctuation is enhanced. In Fig. \ref{figF}, we show in (a1) the generation rates in case of $\gamma_{21} = 0.1 \gamma_{31}$, As we can see that under the relatively larger decoherence rate, $R_{s}^{F}$ deviates from $R_{s}^{F}$ significantly. This results in an approximately 10 times smaller amplitude of the photon correlation (a2) as compared with the result shown in Fig. \ref{figD}.

Another noticeable feature in Fig. \ref{figC} ($N=1.0\, \mu m ^{-3}$) is that $R_{s}^{F}$ and $R_{as}^{F}$ is larger than $R_{s}^{D}$ and $R_{as}^{D}$. The difference between them depends on the atomic density. In Fig. \ref{figF}(b1) we show the generation rate when a lower atomic density ($N=0.5 \,\mu m ^{-3}$) is chosen. The difference between the noise and the deterministic photon pair gets larger, especially when a larger driving field is applied. 
One can attribute this property to the relations between the coefficients in Eqs. (\ref{eq.s}) and (\ref{eq.as}) and the atomic density. $\kappa_s$, $\kappa_as$, $g_R$ and $\Gamma$, see Eq. (\ref{eqG})-(\ref{eqg}) in Appendix \ref{appB} are proportional to $N$, where the coefficient related to the fluctuation $\xi_\alpha^{s/as}$ (listed in Appendix \ref{appB} as well) depends on $\sqrt{N}$. The degree of correlation is dramatically reduced, as shown in (b2). The feature of suppression of noise by increasing atomic density is manifested in the traditionally four-level system as well. In our system with the enhanced generation rate by the coupling field, it is clear that the atomic density should be increased accordingly to suppress the noise.

\section{\label{sec5}Conclusions}

In this paper, we present a theoretical investigation of the coherent control on the SFWM-based biphoton generation by a driving field. Such driving field forms a ladder-style three-level subsystem together with the pumping field. To avoid the dark state, the transition between the ground and Rydberg state is largely detuned from its double-photon resonance.
By assuming the weak coupling effect of the driving and pumping field, and a relatively law atomic density, we investigate the enhancement and control of the photon pairs in a region where the Rydberg atomic correlation is negligible. 
Due to the dynamical shift of the intermediate state, the third-order SFWM nonlinear coefficient is enhanced by the driving field. And the rate of the increment is larger than that of the Raman gain and absorption. Thus the noise of the generated photons is still under control, although it is increased with the driving field as well. 
This leads to an enhanced generation rate with relatively high degree of the correlation.
For an ever higher intensity of the driving, the effects of the fluctuation become severe, then one need a correspondingly higher atomic density to suppress the noise in order to maintain the strong photon correlation.

\section{Acknowledge}
The work is supported by the National Natural Science Foundation of China (12074061).

\appendix
\section{\label{appA}The equations for the element of the density matrix}
Let $\Gamma_{m}$ stands for the total decay rate from state $|m\rangle$, and $\Gamma_{mn}$ is the decay rate from state $|m\rangle$ to state $|n\rangle$.  The corresponding dephasing rate $\gamma_{mn}$ can be written as:
$\gamma_{54}=\gamma_{45}=(\Gamma_{5}+\Gamma_{4})/2$, 
$\gamma_{53}=\gamma_{35}=(\Gamma_{5}+\Gamma_{3})/2$, 
$\gamma_{31}=\gamma_{32}=\Gamma_{3}/2$,
$\gamma_{41}=\gamma_{42}=\Gamma_{4}/2$,
and 
$\gamma_{43}=\gamma_{34}=(\Gamma_{4}+\Gamma_{3})/2$.
Then the elements of the density matrix, based on Eq. (\ref{sigma.mn}) satisfy the following equations.
\begin{widetext}
\begin{subequations}
\begin{equation}\notag
\begin{split}
\partial_t \hat  {\sigma} _{11}= \hat {F}_{11}+\Gamma_{41}\hat  {\sigma} _{44}+\Gamma_{31}\hat  {\sigma} _{33}-i\left(g_{as}\hat {a}_{as}\hat {\sigma} _{31}+\Omega _{p}\hat  {\sigma} _{41}-g_{as}\hat {a}_{as}^ { \dagger }\hat {\sigma} _{13}- \Omega _{p}^* \hat  {\sigma} _{14} \right) ;
\end{split}
\end{equation}
\begin{equation}\notag
\begin{split}
\partial_t \hat{\sigma} _{22}=\hat {F}_{22}+\Gamma_{42}\hat {\sigma} _{44}+\Gamma_{32}\hat  {\sigma} _{33}
-i\left( \Omega _{c}\hat  {\sigma} _{32}+g_{s}\hat {a}_{s}\hat {\sigma} _{42}- \Omega _{c}^*\hat  {\sigma} _{23}-g_{s}\hat {a}_{s}^ { \dagger }\hat {\sigma} _{24}\right);
\end{split}
\end{equation}
\begin{equation}\notag
\begin{split}
\partial_t \hat  {\sigma} _{33}=\hat {F}_{33}-\left(\Gamma_{32}+\Gamma_{31}\right)\hat {\sigma} _{33}+\Gamma_{53}\hat {\sigma} _{55}-i\left(g_{as}\hat {a}_{as}^ { \dagger }\hat {\sigma} _{13}+ \Omega _{c}^*\hat  {\sigma} _{23}-g_{as}\hat {a}_{as}\hat {\sigma} _{31}- \Omega _{c}\hat  {\sigma} _{32} \right);
\end{split}
\end{equation}
\begin{equation}\notag
\begin{split}
\partial_t \hat  {\sigma} _{44}=\hat {F}_{44}-\left(\Gamma_{42}+\Gamma_{41}\right)\hat {\sigma} _{44}+\Gamma_{54}\hat  {\sigma} _{55}-i
\left( \Omega _{p}^*\hat {\sigma} _{14}+\Omega _{d}\hat {\sigma} _{54}+g_{s}\hat {a}_{s}^ { \dagger }\hat {\sigma} _{24}- \Omega _{p}\hat  {\sigma} _{41}- \Omega _{d}^*\hat  {\sigma} _{45}-g_{s}\hat {a}_{s}\hat {\sigma} _{42}\right);
\end{split}
\end{equation}
\begin{equation}\notag
\begin{split}
\partial_t \hat {\sigma} _{55}=&\hat {F}_{55}-\left(\Gamma_{53}+\Gamma_{54}\right)\hat {\sigma} _{55}-i\left( \Omega _{d}^*\hat  {\sigma} _{45}- \Omega _{d}\hat  {\sigma} _{54}\right);
\end{split}
\end{equation}
\begin{equation}\notag
\begin{split}
\partial_t \hat {\sigma} _{54}=\hat {F}_{54}-\left(\gamma_{54}-i\Delta _{d}\right)\hat  {\sigma} _{54}-i\left(\Omega _{d}^*\hat{\sigma} _{44}- \Omega _{p}\hat  {\sigma} _{51}-g_{s}\hat {a}_{s}\hat {\sigma} _{52}- \Omega _{d}^*\hat  {\sigma} _{55}\right);
\end{split}
\end{equation}
\begin{equation}\notag
\begin{split}
\partial_t \hat  {\sigma} _{51}=\hat {F}_{51}-\left(\gamma_{51}-i\Delta _{15}\right)\hat  {\sigma} _{51}-i\left( \Omega _{d}^*\hat  {\sigma} _{41}-g_{as}\hat {a}_{as}^ { \dagger }\hat {\sigma} _{53}- \Omega _{p}^*\hat  {\sigma}_{54}\right);
\end{split}
\end{equation}
\begin{equation}\notag
\begin{split}
\partial_t \hat  {\sigma} _{41}=\hat {F}_{41}-\left(\gamma_{41}-i\Delta_{p}\right)\hat  {\sigma} _{41}-i \left(\Omega _{p}^*\hat  {\sigma} _{11}+g_{s}\hat {a}_{s}^ { \dagger }\hat{\sigma} _{21}+\Omega _{d}\hat {\sigma} _{51}-g_{as}\hat {a}_{as}^ { \dagger }\hat {\sigma} _{43}- \Omega _{p}^*\hat  {\sigma} _{44}\right);
\end{split}
\end{equation}
\begin{equation}\notag
\begin{split}
\partial_t \hat  {\sigma} _{32}=\hat {F}_{32}-\left(\gamma_{32}-i\Delta_{c}\right)\hat {\sigma} _{32}-i \left( g_{as}\hat {a}_{as}^ { \dagger }\hat {\sigma} _{12}+\Omega _{c}^*\hat  {\sigma} _{22}-\Omega _{c}^*\hat {\sigma} _{33}-g_{s}\hat {a}_{s}^ { \dagger }\hat {\sigma} _{34} \right);
\end{split}
\end{equation}
\begin{equation}\notag
\begin{split}
\partial_t \hat  {\sigma} _{25}=\hat {F}_{25}-\left[\gamma_{25}+i\left(\Delta _{15}+\omega\right)\right]\hat  {\sigma} _{25}-i\left (\Omega _{c}\hat {\sigma} _{35}+g_{s}\hat {a}_{s}\hat{\sigma} _{45}- \Omega _{d}\hat  {\sigma} _{24}\right  );
\end{split}
\end{equation}
\begin{equation}\notag
\begin{split}
\partial_t \hat  {\sigma} _{35}=\hat {F}_{35}-\left[\gamma_{35}+i\left(\Delta _{15}+\omega-\Delta _{c}\right)\right]\hat  {\sigma} _{35}-i\left (g_{as}\hat {a}_{as}^ { \dagger }\hat {\sigma} _{15}+\Omega _{c}^*\hat  {\sigma} _{25}- \Omega _{d}\hat  {\sigma} _{34}\right );
\end{split}
\end{equation}
\begin{equation}\notag
\begin{split}
\partial_t \hat  {\sigma} _{24}=\hat {F}_{24}-\left[\gamma_{24}+i\left(\Delta _{p}+\omega\right)\right]\hat {\sigma} _{24}-i \left [\Omega _{c}\hat  {\sigma} _{34}+g_{s}\hat {a}_{s}\left(\hat {\sigma} _{44}-\hat {\sigma} _{22}\right)- \Omega _{p}\hat  {\sigma} _{21}- \Omega _{d}^*\hat  {\sigma} _{25}\right ];
\end{split}
\end{equation}
\begin{equation}\notag
\begin{split}
\partial_t \hat  {\sigma} _{34}=\hat {F}_{34}-\left[\gamma_{34}+i\left(\Delta _{p}+\omega-\Delta _{c}\right)\hat {\sigma} _{34}\right]-i \left (g_{as}\hat {a}_{as}^ { \dagger }\hat {\sigma} _{14}+ \Omega _{c}^*\hat {\sigma} _{24}- \Omega _{p}\hat  {\sigma} _{31}- \Omega _{d}^*\hat {\sigma} _{35}-g_{s}\hat {a}_{s}\hat {\sigma} _{32}\right);
\end{split}
\end{equation}
\begin{equation}\notag
\begin{split}
\partial_t \hat  {\sigma} _{21}=\hat {F}_{21}-\left(\gamma_{21}+i\omega\right)\hat  {\sigma} _{21}-i\left ( \Omega _{c}\hat  {\sigma} _{31}+g_{s}\hat {a}_{s}\hat{\sigma} _{41}-g_{as}\hat {a}_{as}^ { \dagger }\hat {\sigma} _{23}- \Omega _{p}^*\hat  {\sigma} _{24}\right );
\end{split}
\end{equation}
\begin{equation}\notag
\begin{split}
\partial_t \hat  {\sigma} _{31}=\hat {F}_{31}-\left[\gamma_{31}+i\left(\omega-\Delta _{c}\right)\hat  {\sigma} _{31}\right]-i\left [g_{as}\hat {a}_{as}^ { \dagger }\left(\hat {\sigma} _{11}-\hat {\sigma} _{33}\right)+\Omega _{c}^*\hat  {\sigma} _{21}-\Omega _{p}^*\hat  {\sigma} _{34}\right ].
\end{split}
\end{equation}
\end{subequations}

\section{\label{appB} Coefficients in the Heisenberg-Langevin equations}
The coefficients appear in the Eq. (\ref{eq.s}) and (\ref{eq.as}) are quite complex. However, they can be written in relatively compact forms if expressed in terms of the zeroth-order solutions.
\begin{equation}\label{eqG}
\Gamma  ={-i \mathcal{N} \sigma  \gamma _{31}\Omega^{-1} }\Big(\mathscr{A}_1 \langle\sigma_{22}^{(0)}\rangle+\mathscr{A}_2 \langle\sigma_{32}^{(0)}\rangle+\mathscr{A}_5 \langle\sigma_{41}^{(0)}\rangle-\mathscr{A}_1 \langle\sigma_{44}^{(0)}\rangle+\mathscr{A}_3 \langle\sigma_{45}^{(0)}\rangle\Big);
\end{equation}
\begin{equation}
\kappa _{s}={-i \mathcal{N} \sigma  \gamma _{31}\Omega^{-1} }\Big(\mathscr{B}_1\langle\sigma_{22}^{(0)}\rangle+ \mathscr{B}_3\langle\sigma_{32}^{(0)}\rangle+\mathscr{B}_5\langle\sigma_{41}^{(0)}\rangle- \mathscr{B}_1\langle\sigma_{44}^{(0)}\rangle+ \mathscr{B}_4\langle\sigma_{45}^{(0)}\rangle\Big);
\end{equation}
\begin{equation}
\kappa _{as}={-i \mathcal{N} \sigma  \gamma _{31}\Omega^{-1} }\Big(\mathscr{A}_6 \langle\sigma_{11}^{(0)}\rangle-\mathscr{A}_2 \langle\sigma_{14}^{(0)}\rangle+\mathscr{A}_4 \langle\sigma_{15}^{(0)}\rangle-\mathscr{A}_5 \langle\sigma_{23}^{(0)}\rangle-\mathscr{A}_6 \langle\sigma_{33}^{(0)}\rangle\Big);
\end{equation}
\begin{equation}\label{eqg}
g_{R}={-i \mathcal{N} \sigma  \gamma _{31}\Omega^{-1} }\Big(\mathscr{B}_6\langle\sigma_{11}^{(0)}\rangle- \mathscr{B}_3\langle\sigma_{14}^{(0)}\rangle+ \mathscr{B}_2\langle\sigma_{15}^{(0)}\rangle- \mathscr{B}_5\langle\sigma_{23}^{(0)}\rangle- \mathscr{B}_6\langle\sigma_{33}^{(0)}\rangle\Big).
\end{equation}
Where $\mathcal{N} = N / V$ is the atomic density, $\sigma$ is the absorption cross section of the atom.
We define complex quantities $g_1=\gamma _{21}+i \omega$, $g_2=\gamma _{24}+i \left(\omega +\Delta _p\right)$, $g_3=\gamma _{25}+i \left(\omega +\Delta _{15}\right)$ 
$g_4=\gamma _{31}+i \left(\omega -\Delta _c\right)$, $g_5=\gamma _{34}+i \left(-\Delta _c+\omega +\Delta _p\right)$, $g_6=\gamma _{35}+i \left(-\Delta _c+\omega +\Delta _{15}\right)$
to collect decoherence rate and detunings together. After introducing a symbol to represent the product of those quantities  $g_{\alpha\beta\gamma\cdots} = g_\alpha g_\beta g_\gamma \cdots$, then $\Omega$, $\mathscr{A}_\alpha$, and $\mathscr{B}_\alpha$ in the above equations can be expressed as:
$\Omega =g_{23} |\Omega _c \Omega _d|^2-2 |\Omega _c|^4 |\Omega _d|^2+g_{25} |\Omega _c|^4+g_{2356} |\Omega _c|^2$, 
$\mathscr{A}_1=i g_3 |\Omega _c\Omega _d|^2+i g_5 |\Omega _c|^4+i g_{356}| \Omega _c|^2,\mathscr{A}_2=-\Omega _c| \Omega _c\Omega _d|^2+g_{36}\Omega _c| \Omega _c|^2+|\Omega _c|^4\Omega _c$, 
$\mathscr{A}_3=g_{56} |\Omega _c|^2 \Omega _d-|\Omega _c|^4 \Omega _d+|\Omega _c \Omega _d|^2\Omega _d,\mathscr{A}_4=-i g_3 |\Omega _c|^2\Omega _c \Omega _d-i g_5  |\Omega _c|^2\Omega _c \Omega _d$,
$\mathscr{A}_5=|\Omega _c \Omega _d|^2 \Omega _p-g_{36} |\Omega _c|^6| \Omega _p|^2+g_{3456} \Omega _p,\mathscr{A}_6=-i g_3 \Omega _c |\Omega _d|^2 \Omega _p-i g_5 |\Omega _c|^2\Omega _c \Omega _p-i g_{356} \Omega _c \Omega _p$,
$\mathscr{B}_1=i g_3 \Omega _c \Omega _d^2 \Omega _p+i g_5 |\Omega _c|^2\Omega _c \Omega _p+i g_{356} \Omega _c \Omega _p,\mathscr{B}_2=-i g_3 |\Omega _c|^2 \Omega _d \Omega _p-i g_5 |\Omega _c|^2 \Omega _d \Omega _p$,
$\mathscr{B}_3=-|\Omega _c\Omega _d|^2 \Omega _p+g_{36} |\Omega _c|^2 \Omega _p+|\Omega _c|^4 \Omega _p,\mathscr{B}_4=g_{56} \Omega _c \Omega _d \Omega _p$,
$\mathscr{B}_5=-g_{23} \Omega _c |\Omega _d|^2-g_{25} |\Omega _c|^2\Omega _c-g_{2356} \Omega _c,\mathscr{B}_6=-i g_{356} |\Omega _p|^2$,

The coefficient of the noise operators are:
\begin{center}
  \renewcommand{\arraystretch}{2.0}
  \begin{tabular}{l  l }
  $\xi _{21}^{\text{as}}=2 i \sqrt{\mathcal{N} \sigma  \gamma _{31}}{\Omega^{-1} } (-i g_{56} \Omega _c |\Omega _d|^2-i g_{25} |\Omega _c|^2\Omega _c-i g_{2356} \Omega _c)\;$
  & $\xi _{31}^{\text{as}}=2 i \sqrt{\mathcal{N} \sigma  \gamma _{31}}{\Omega^{-1} } ( g_5 |\Omega _c \Omega _p|^2+g_{356} |\Omega _p|^2)$\\
  $\xi _{24}^{\text{as}}=2 i \sqrt{\mathcal{N} \sigma  \gamma _{31}}{\Omega^{-1} } (g_5 |\Omega _c|^2\Omega _c \Omega _p+g_{356} \Omega _c \Omega _p),$
  & $\xi _{34}^{\text{as}}=2 i \sqrt{\mathcal{N} \sigma  \gamma _{31}} {\Omega^{-1} } (i |\Omega _c \Omega _d|^2 \Omega _p-i g_{36} |\Omega _c|^2 \Omega _p)$\\
  $\xi _{25}^{\text{as}}=2 i \sqrt{\mathcal{N} \sigma  \gamma _{31}} {\Omega^{-1} } (i g_{56} \Omega _c \Omega _d \Omega _p+i \Omega _c |\Omega _d|^2\Omega _d \Omega _p),$
  & $\xi _{35}^{\text{as}}=2 i \sqrt{\mathcal{N} \sigma  \gamma _{31}} {\Omega^{-1} } (g_3 |\Omega _c|^2 \Omega _d \Omega _p+g_5 |\Omega _c|^2 \Omega _d \Omega _p)$ \\
  $\xi _{21}^s=2 i \sqrt{\mathcal{N} \sigma  \gamma _{31}} {\Omega^{-1} } (i g_{3456} \Omega _p-i g_{36} |\Omega _c|^2 \Omega _p),$
  & $\xi _{31}^s=2 i \sqrt{\mathcal{N} \sigma  \gamma _{31}}{\Omega^{-1} } (g_3 \Omega _c |\Omega _d|^2 \Omega _p+g_{356} \Omega _c \Omega _p)$\\
  $\xi _{24}^s=2 i \sqrt{\mathcal{N} \sigma  \gamma _{31}} {\Omega^{-1} } (g_3 |\Omega _c \Omega _d|^2+g_{356} |\Omega _c|^2),$
  &$\xi _{34}^s=2 i \sqrt{\mathcal{N} \sigma  \gamma _{31}} {\Omega^{-1} } (i |\Omega _c|^2\Omega _c \Omega _d^2-i g_{36}|\Omega _c|^2\Omega _c)$\\
  $\xi _{25}^s=2 i \sqrt{\mathcal{N} \sigma  \gamma _{31}} {\Omega^{-1} } (i g_{56} |\Omega _c|^2 \Omega _d+i |\Omega _c|^2 |\Omega _d|^2\Omega _d),$ 
  & $\xi _{35}^s=2 i \sqrt{\mathcal{N} \sigma  \gamma _{31}}  {\Omega^{-1} } (g_5 |\Omega _c|^2\Omega _c \Omega _d+g_3 |\Omega _c|^2\Omega _c \Omega _d)$
   \end{tabular}
  \end{center}

\section{\label{appC}Diffusion Coefficients}
The collective Langevin noise operators are defined, in the similar way of the collective slowly varying atomic operators, as
\begin{equation}
\hat{F}_{mn}(z,t)=\frac{1}{N}\sum_{i\in N}\hat{F}_{mn}^{[i]}(t),
\end{equation}
where $\hat{F}_{mn}^{[i]}(t)$ is Langevin noise operator for the $i^\text{th}$ atom. $\hat{F}_{mn}(z,t)$ is subject to the correlation of
\begin{equation}
\langle\hat{F}_{mn}(t,z)\hat{F}_{m'n'}(t',z')\rangle=\frac{L}{N}\mathcal{D}_{mn,m'n'}(t,z)\delta(t-t')\delta(z-z').
\end{equation}
In the frequency domain the noise correlation is
\begin{equation}
\langle\hat{F}_{mn}(\omega,z)\hat{F}_{m'n'}(\omega',z')\rangle=\frac{L}{2\pi N}\mathcal{D}_{mn,m'n'}\delta(\omega-\omega')\delta(z-z').
\end{equation}
The diffusion coefficients $\mathcal{D}_{mn,m'n'}$, based on the generalized fluctuation-dissipation theorem and Einstein relation, can be obtained from the relation:
\begin{equation}
\begin{split}
\mathcal{D}_{mn,m'n'}&=\mathscr{D}(\hat{\sigma}_{mn}\hat{\sigma}_{m'n'}) -  \mathscr{D}(\hat{\sigma}_{mn})\hat{\sigma}_{m'n'} \\
&- \hat{\sigma}_{mn} \mathscr{D}(\hat{\sigma}_{m'n'}).
\end{split}
\end{equation}
The notation $\mathscr{D}(\hat{\sigma}_{mn})$ denotes the deterministic part of
the Heisenberg equation of motion for $\hat{\sigma}_{mn}$ (Equations in Appendix \ref{appA} without the fluctuation operators).  Then those for the noise of Stokes photons are


  \begin{center}
  \renewcommand{\arraystretch}{1.5}
  \begin{tabular}{l  l  l }
    $\mathcal{D}_{12,24}=\gamma_{12}\langle\sigma_{14}^{(0)}\rangle\quad\quad$ 
    & $\mathcal{D}_{12,25}=\gamma_{12}\langle\sigma_{15}^{(0)}\rangle\quad\quad$ 
    & $\mathcal{D}_{13,31}=\Gamma_{41}\langle\sigma_{44}^{(0)}\rangle+\Gamma_{31}\langle\sigma_{33}^{(0)}\rangle+(\Gamma_{31}+\Gamma_{32})\langle\sigma_{11}^{(0)}\rangle$ \\
    $\mathcal{D}_{13,34}=(\Gamma_{31}+\Gamma_{32})\langle\sigma_{14}^{(0)}\rangle\quad\quad$ 
    &$\mathcal{D}_{13,35}=\Gamma_{31}+\Gamma_{32}\langle\sigma_{15}^{(0)}\rangle\quad\quad$ 
    &$\mathcal{D}_{42,21}=\gamma_{21}\langle\sigma_{41}^{(0)}\rangle$ \\
    $\mathcal{D}_{42,24}=\Gamma_{54}\langle\sigma_{55}^{(0)}\rangle\quad\quad$ 
    & $\mathcal{D}_{43,31}=(\Gamma_{31}+\Gamma_{32})\langle\sigma_{41}^{(0)}\rangle\quad\quad$ 
    & $\mathcal{D}_{43,34}=\Gamma_{54}\langle\sigma_{55}^{(0)}\rangle+(\Gamma_{31}+\Gamma_{32})\langle\sigma_{44}^{(0)}\rangle$ \\
    $\mathcal{D}_{43,35}=(\Gamma_{31}+\Gamma_{32})\langle\sigma_{45}^{(0)}\rangle\quad\quad$
    & $\mathcal{D}_{52,21}=\gamma_{21}\langle\sigma_{51}^{(0)}\rangle\quad\quad$ 
    & $\mathcal{D}_{53,31}=(\Gamma_{31}+\Gamma_{32})\langle\sigma_{51}^{(0)}\rangle$\\
    $\mathcal{D}_{53,34}=(\Gamma_{31}+\Gamma_{32})\langle\sigma_{54}^{(0)}\rangle\quad\quad$
    & $\mathcal{D}_{53,35}=(\Gamma_{31}+\Gamma_{32})\langle\sigma_{55}^{(0)}\rangle\quad\quad$
  \end{tabular}
  \end{center}
And those for the anti-Stokes photons are
  \begin{center}
  \renewcommand{\arraystretch}{1.5}
  \begin{tabular}{l  l }
    $\mathcal{D}_{21,12}=\Gamma_{32}\langle\sigma_{33}^{(0)}\rangle+\Gamma_{42}\langle\sigma_{44}^{(0)}\rangle+2\gamma_{21}\langle\sigma_{22}^{(0)}\rangle$ 
    & $\mathcal{D}_{21,13}=\gamma_{21}\langle\sigma_{23}^{(0)}\rangle$ \\
    $\mathcal{D}_{31,12}=\gamma_{12}\langle\sigma_{32}^{(0)}\rangle$
    & $\mathcal{D}_{31,13}=\Gamma_{53}\langle\sigma_{55}^{(0)}\rangle$ \\
    $\mathcal{D}_{24,42}=\gamma_{32}\langle\sigma_{33}^{(0)}\rangle+\gamma_{42}\langle\sigma_{44}^{(0)}\rangle+(\Gamma_{41}+\Gamma_{42}) \langle\sigma_{22}^{(0)}\rangle$ 
    & $\mathcal{D}_{24,43}=(\Gamma_{41}+\Gamma_{42}) \langle\sigma_{23}^{(0)}\rangle$ \\
    $\mathcal{D}_{25,52}=\Gamma_{32}\langle\sigma_{33}^{(0)}\rangle+\Gamma_{42}\langle\sigma_{44}^{(0)}\rangle+(\Gamma_{53}+\Gamma_{54})\langle\sigma_{22}^{(0)}\rangle\quad\quad$ 
    & $\mathcal{D}_{25,53}=(\Gamma_{53}+\Gamma_{54})\langle\sigma_{23}^{(0)}\rangle$ \\
    $\mathcal{D}_{35,52}=(\Gamma_{53}+\Gamma_{54})\langle\sigma_{32}^{(0)}\rangle$
    &$\mathcal{D}_{35,53}=\Gamma_{53}\langle\sigma_{55}^{(0)}\rangle+(\Gamma_{53}+\Gamma_{54})\langle\sigma_{33}^{(0)}\rangle$\\
    $\mathcal{D}_{34,43}=\Gamma_{53}\langle\sigma_{55}^{(0)}\rangle+(\Gamma_{41}+\Gamma_{42}) \langle\sigma_{33}^{(0)}\rangle$ 
    & $\mathcal{D}_{34,42}=(\Gamma_{41}+\Gamma_{42}) \langle\sigma_{32}^{(0)}\rangle$
  \end{tabular}
  \end{center}

\end{widetext}



%

\end{document}